\documentclass[a4paper]{PoS}

\title{CMS software and computing for LHC Run 2}

\ShortTitle{CMS software and computing for LHC Run 2}

\author{\speaker{Kenneth Bloom}\thanks{For the CMS Collaboration.  This presentation is similar to one I gave at the DPF 2015 conference, and thus there is likely some textual overlap with K.~Bloom, arXiv:1509.08180.}\\
      University of Nebraska-Lincoln\\
      E-mail: \email{kenbloom@unl.edu}}

\abstract{
The CMS offline software and computing system has successfully met the challenge of LHC Run~2. In this presentation, we will discuss how the entire system was improved in anticipation of increased trigger output rate, increased rate of pileup interactions and the evolution of computing technology. The primary goals behind these changes was to increase the flexibility of computing facilities where ever possible, as to increase our operational efficiency, and to decrease the computing resources needed to accomplish the primary offline computing workflows. These changes have resulted in a new approach to distributed computing in CMS for Run~2 and for the future as the LHC luminosity should continue to increase. We will discuss changes and plans to our data federation, which was one of the key changes towards a more flexible computing model for Run~2. Our software framework and algorithms also underwent significant changes. We will summarize the our experience with a new multi-threaded framework as deployed on our prompt reconstruction farm for 2015 and across the CMS WLCG Tier-1 facilities. We will discuss our experience with a analysis data format which is ten times smaller than our primary Run~1 format. This ``mini-AOD'' format has proven to be easier to analyze while be extremely flexible for analysts. Finally, we describe improvements to our workflow management system that have resulted in increased automation and reliability for all facets of CMS production and user analysis operations.
}

\FullConference{38th International Conference on High Energy Physics\\
		  3-10 August 2016\\
		 Chicago, USA}

\begin{document}
July 4, 2012 was a happy day for all of particle physics, but Joe Incandela, the spokesperson of the Compact Muon Solenoid (CMS) Collaboration, was wearing a particularly large smile.  Why?  It was because CMS software and computing had enabled the discovery of the Higgs boson.  CMS had used every drop of data available from the Large Hadron Collider (LHC) for searches in all five main Higgs decay channels, and had the necessary simulation samples to complete measurements.  This was not the case for the competitor experiment, even though it had greater computing resources~\cite{bib:REBUS}.  But the competitor has hardly stood still since 2012, and neither has CMS.  Here we describe the innovations in CMS software and computing that will make the experiment successful again in Run~2.

Run~2, and the 2016 running year in particular, have presented significant challenges to CMS.  CMS is exploring a new energy domain, with opportunities for early discovery.  The LHC has provided its highest luminosities ever, and the data has arrived quickly, so all systems must be prepared.  In addition, the computing requirements are substantially larger than those of Run~1.  The event rate to storage is 1~kHz or more, a factor of 2.5 greater than in Run~1, and the pileup rate is reaching 50 interactions per beam crossing.  Without any improvements to the software, a factor of six increase in CPU would be necessary to reconstruct the data in a timely fashion, which would be challenging to achieve within current budgets.  To address this challenge, CMS used the long shutdown to modernize its software and computing, delivering a system of increased agility and flexibility that enables physics discoveries that is built on the extremely successful systems of Run~1.

\begin{figure}
  \includegraphics[width=.6\textwidth]{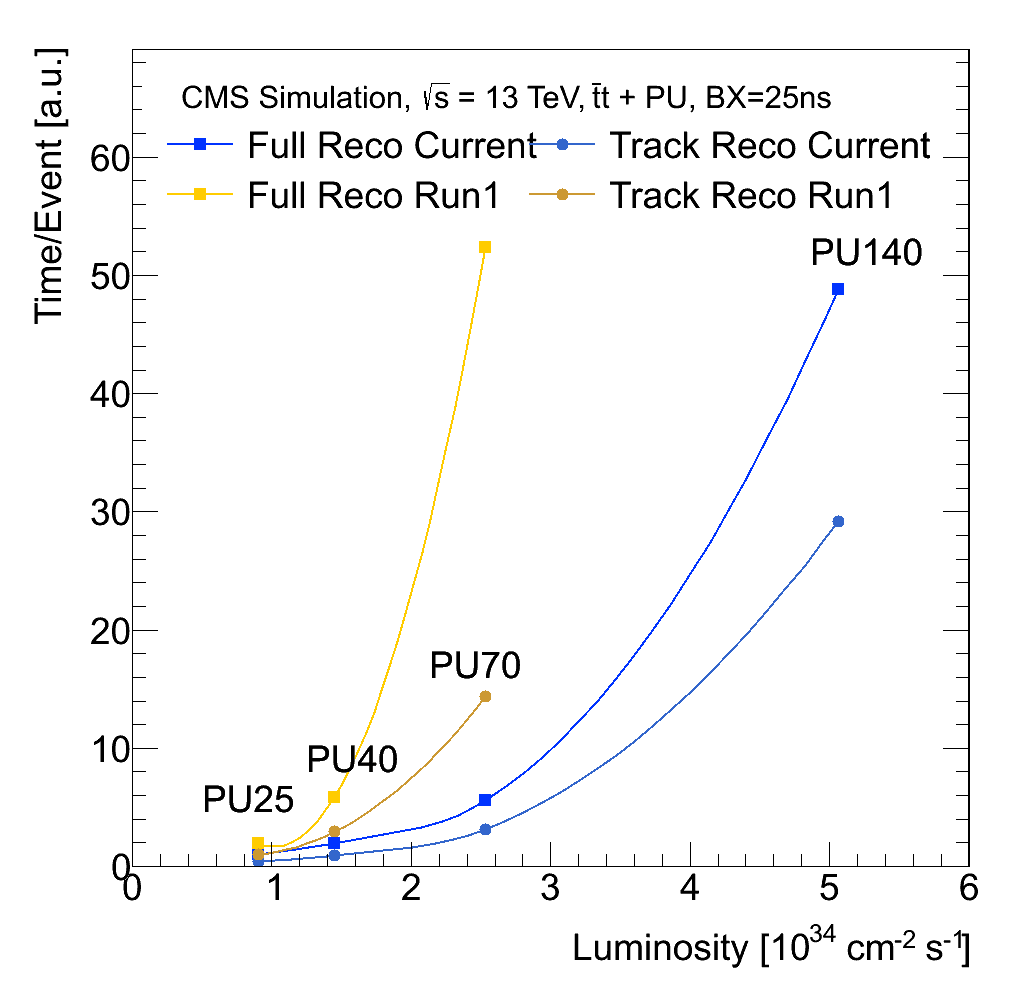}
\centering
  \caption{Reconstruction time for a $t\bar{t}$ event as a function of instantaneous luminosity for two different versions of CMS reconstruction software. The time for track reconstruction only is broken out separately.}
     \label{fig:recotime}
     \end{figure}

Key improvements were made in the performance of the reconstruction software, which dominates CMS CPU usage.  As shown in Figure~\ref{fig:recotime}, CMS has reduced the event reconstruction time, and in particular has decreased the sensitivity to pileup.  The most important improvements were in track reconstruction, the leading component of reconstruction time.  Some of the improvements came from technical changes to the code, while others came from new algorithms that reduced the number of fake tracks and thus sped execution.  The simulation time was also improved by reducing the time that {\tt GEANT4} spent tracking low-energy particles.

\begin{figure}
  \includegraphics[width=.6\textwidth]{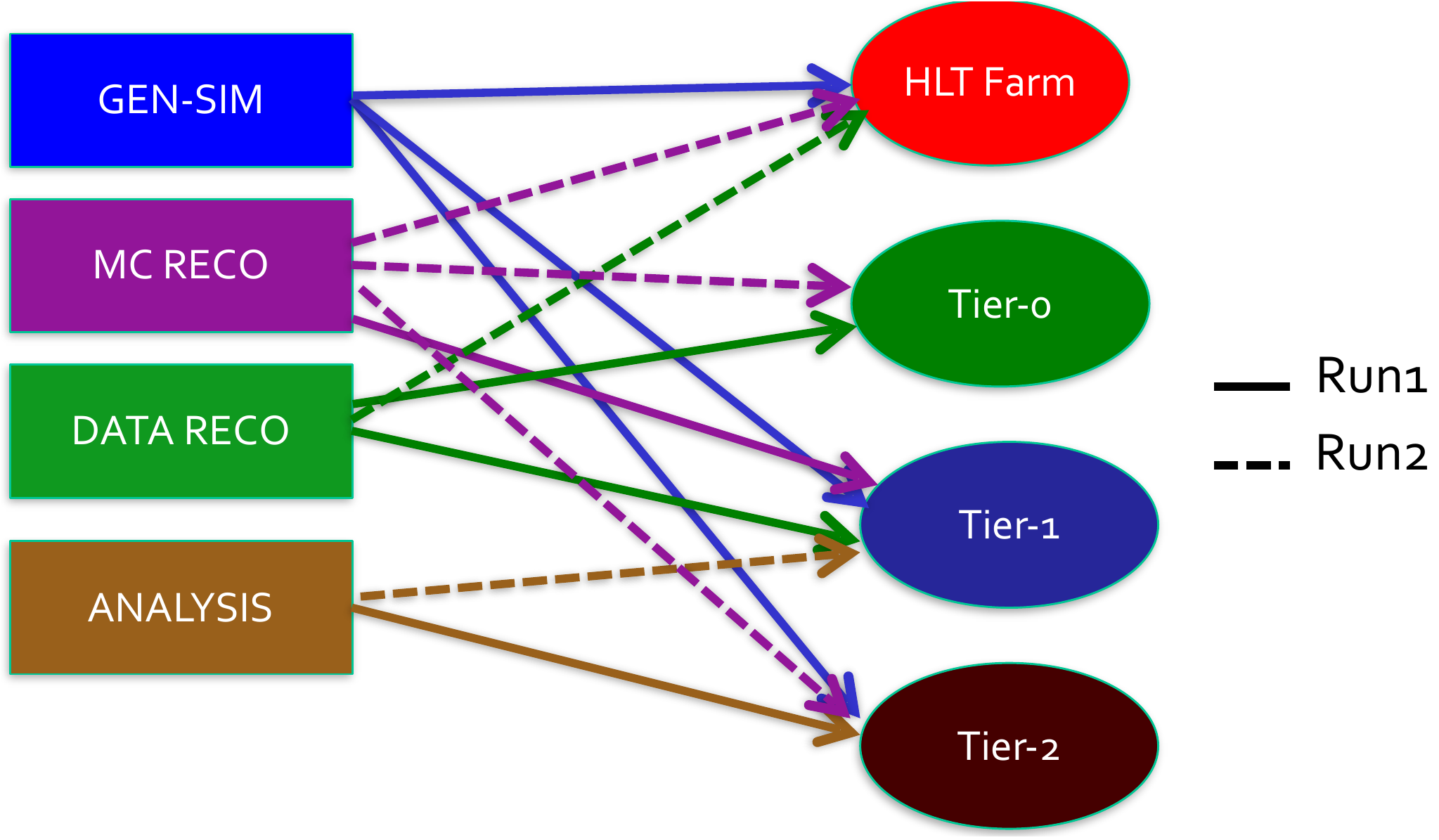}
\centering
\caption{Various workflows (boxes) were run at certain facilities (ovals) during Run~1, as indicated by the solid arrows.  The dashed arrows indicate the additional workflows-to-facilities mappings that are possible in Run~2.}
     \label{fig:workflows}
     \end{figure}

CMS also undertook a major effort to use its computing facilities in more flexible and heterogeneous ways.   Figure~\ref{fig:workflows} indicates the principal workflows of CMS computing, and which facilities execute them.  In Run~2, a greater diversity of workflows run at the sites.  In particular, the high level trigger (HLT) farm can be used for organized processing during technical stops, Tier-2 sites are used for reconstruction tasks that were previously only done at Tier-1 sites, and user analysis jobs are executed at additional sites.  The more places where work can run, the faster the work ultimately goes.

     The increased diversity in site usage is supported by new services.  The ``Any Data, Anytime, Anywhere'' (AAA) data federation allows CMS applications to read data efficiently over wide-area networks, thus relaxing the constraints on the relative locations of data and workflows.~\cite{bib:AAA}.  Disk-tape separation at Tier-1 sites has allowed greater control over what datasets are available on disk, and in addition allows the disk-resident data to be used in workflows anywhere in the world through the AAA system.  A new dynamic data management system automatically places datasets at computing sites when they are created and then deletes them when they are no longer needed, based on ``popularity'' information, for more agile and efficient use of disk space.  There is now a global job pool for resource provisioning through the glideinWMS system.~\cite{bib:glideinWMS}.  This allows for central control of job priorities, and brings a simplified infrastructure compared to running different job pools for different tasks.  It scales sufficiently to operate Tier-1, Tier-2 and opportunistic resources in the single pool.  glideinWMS can provision cloud infrastructures, which allows the use of the HLT and opportunistic and commercial clouds.  This gives CMS the capability to straightforwardly burst into extra resources if necessary.  Finally, the establishment of a 100~Gbps transatlantic network link by ESnet facilities interactions between CERN and the Western Hemisphere.

Some technical innovations in the software framework have also brought greater efficiencies and new opportunities.  The code for both the simulation and reconstruction executables is now multi-threaded.  It can use several CPU cores concurrently to reconstruct multiple events simultaneously.  This reduces the demands on the computing infrastructure, as there are fewer open files and fewer jobs overall.  The time to process a luminosity block (the quantum of CMS data) is reduced, as is needed for the higher trigger rates.  There is also a huge reduction in the memory required per CPU core with little loss in throughput.  The advent of the multi-threaded framework could allow the experiment to explore novel processing architectures.  It also enables the use of multi-core pilot jobs that do their own internal dynamic partitioning of resources for greater efficiency.

CMS has also made physics analysis easier, more flexible and less resource-intensive through improved user tools.  {\tt CRAB3}, the new analysis job submission tool, has many user-friendly features.  Jobs that fail are automatically retried, and job-tracking tools are improved.  The delivery of output files has become more reliable thanks to centralized handling of data transfers.  The client is thinner than before; moving more of the logic to the server side has allowed for easier upgrades.  {\tt CRAB3} also fully exploits the features of HTCondor~\cite{bib:condor} and glideinWMS, including the overflowing of jobs from busy to less-busy sites.  Most users now analyze the {\tt miniAOD} format, which at 30~kB/event is one tenth the size of the {\tt AOD} event format that was used for most Run~1 analyses.  The new format is designed to serve about 80\% of analyses, and the smaller size makes it easier to keep more of the data at desired processing locations.  In addition, the AAA data federation supports users by making job location independent of data location.  This is a major enabler for data analysis performed at universities, which might not have the resources to support a large storage system that can hold all the data needed for a CMS analysis.

A very interesting new frontier for CMS computing is the use of dynamically-provisioned resources.  An ability to rapidly expand resources for burst needs could be a game changer for resource provisioning.  If it is successful (which is admittedly a significant ``if''), then it would be possible to purchase processing resources only for average needs, rather than for the peaks.  There has been a very successful demonstration of dynamic provisioning of Amazon Web Services (AWS) resources via the Fermilab HEPCloud, as described elsewhere at this conference.~\cite{bib:HEPCloud}  The demonstration had great diversity of usage for the experiment; all types of CMS workflows were performed in all AWS resource instances in all AWS availability zones.  It achieved the scale goals of running at least 50,000 jobs simultaneously, with only 9.5\% ``badput'' and 87\% CPU efficiency.  CMS gained knowledge of how to optimize the cost per unit output.  And most importantly, physics came out of it: 518M events that were generated in early February 2016 were used in analyses that were shown at conferences in March.  CMS now plans for follow-up exercises on all fronts, such as other commercial providers, opportunistic resources on the Open Science Grid, and U.S. national resources such as NERSC and XSEDE.

All of the efforts to prepare for Run~2 have paid off for CMS in good performance of the software and computing systems.  The Tier-0 facility was kept busy through the 2016 run, with the number of job slots in use regularly reaching 20,000.  The actual number of jobs was much less than this thanks to multi-core processing.  The excellent LHC performance meant that much more data had to be transferred out of CERN to Tier-1 sites.  Much work was done in the weeks leading up to this conference to improve the transfer rates.  The grid sites were also very active, with Tier-1 sites typically contributing 30,000 running job cores and Tier-2 sites contributing in excess of 60,000.  The sites performed a mix of activities. Tier-1 sites mostly ran centrally-organized production jobs, but also a small amount of user analysis.  Tier-2 sites ran about an equal mix of production and analysis jobs, but the production jobs included the DIGI-RECO workflow and not just GEN-SIM.  HLT usage has demonstrated sufficient scaling ability for that system.  All of these activities demonstrate that the goal of running more workflows in more places has been met.

The various new services are also running successfully.  The dynamic data management system is ensuring that highly-requested datasets have the most copies available in the distributed computing system.  About 20\% of data reads are done through the data federation, with very low failure rates.  The global job pool is operating at a scale of 150,000 cores, and most are being used in multicore pilots.  Almost all users have transitioned to the {\tt CRAB3} job submission tool.  But most importantly, the CMS software and computing systems have been used for physics results.  Billions of simulated events were produced in under three months for analyses targeted towards this conference.  The last data for those analyses were recorded 19 days before the conference began; all of the data were successfully ingested.  About 40 results on the full data sample were shown at the conference, demonstrating that CMS computing has the necessary throughput for quick turnaround.

CMS software and computing was very successful in Run~1, but could not -- and did not -- stand still for Run~2.  There were significant evolutionary changes to the Run~1 system, resulting in more flexible and more efficient resource usage and better tools for physics users, all of which took advantage of a variety of technical developments.  These changes are now fully operational for Run 2 data taking and analysis, enabling the production of frontier physics results with fast turnaround, even with a harsher experimental environment.  If Nature cooperates, CMS software and computing will have everyone smiling again.

\acknowledgments
I thank my CMS collaborators around the world who have worked to make the software and computing systems successful, and Daniele Bonacorsi, Oliver Gutsche, David Lange and Liz Sexton-Kennedy for their comments on this presentation.  I also thank the ICHEP 2016 organizers for making this conference such a success.

\end{document}